\documentclass[a4paper,10pt]{article}

\usepackage{color}
\usepackage[usenames,dvipsnames]{xcolor} 
\definecolor{DarkGreen}{RGB}{0,64,0}

\usepackage[]{array}
\usepackage[]{amsmath}
\usepackage[]{amssymb}
\usepackage[]{mathrsfs}
\usepackage[]{tensor}
\usepackage[symbol]{footmisc}
\usepackage{bm}

\newcommand{\tr}{\textrm{tr}}

\newcommand{\be}{\begin{eqnarray}}
\newcommand{\ee}{\end{eqnarray}}

\newcommand{\Gam}{\mathbf{\Gamma}}
\newcommand{\R}{\mathbf{R}}

\newcommand{\LL}{\mathbf{L}}

\newcommand{\Th}{\mathbf{\Theta}}
\newcommand{\CS}{\mathbf{\Upsilon}}

\begin{document}
\begin{flushright}
{SISSA 18/2013/FISI \\ TZF-2013-02}
\end{flushright}
\vskip 1cm
\begin{center}
{\LARGE{\bf Symmetries and gravitational Chern-Simons\\[2mm] Lagrangian terms}}
\vskip 1cm

{\Large L.~Bonora$^a$, M.~Cvitan$^b$, P.~Dominis Prester$^c$, S.~Pallua$^b$, I.~Smoli\'c$^b$}\\
{}~\\
\quad \\
{\em ~$~^{a}$International School for Advanced Studies (SISSA/ISAS),}\\
{\em Via Bonomea 265, 34136 Trieste, Italy}
{}~\\
\quad \\
{\em ~$~^{b}$Physics Department, Faculty of Science,}\\
{\em University of Zagreb, p.p.~331, HR-10002 Zagreb, Croatia}
 {}~\\
\quad \\
{\em ~$~^{c}$ Department of Physics, University of Rijeka,}\\
{\em  Radmile Matej\v{c}i\'{c} 2, HR-51000 Rijeka, Croatia}\\
\vskip 1cm
Email: bonora@sissa.it, mcvitan@phy.hr, pprester@phy.uniri.hr, pallua@phy.hr, ismolic@phy.hr

\end{center}

\vskip 2cm 

{\bf Abstract.} 
We consider some general consequences of adding pure gravitational 
Chern-Simons term to manifestly diff-covariant theories of gravity. Extending the result of a previous paper we enlarge the class of metrics for which the inclusion of a gCS term in the action does not affect solutions and corresponding physical quantities. In the case in which such solutions describe black holes (of general horizon topology) we show that the black hole entropy is also unchanged. We arrive at these conclusions by proving three general theorems and studying their consequences. One of the theorems states that the contribution of the gravitational Chern-Simons to the black hole entropy is invariant under local rescaling of the metric.  
\vskip 1cm 

 
\vfill\eject

\section{Introduction}
\label{sec:intro}

This paper is a follow up of previous papers in which we have analyzed the consequences of adding a purely gravitational Chern-Simons (gCS) term \cite{CS} to a manifestly diffeomorphism invariant gravitational action in $(4k-1)$-dimensional spacetime, which is a generalization of the idea originally introduced in $D=3$ dimensions in \cite{DJT1,DJT2}. Following a proposal by Tachikawa \cite{Tachikawa:2006sz}, in \cite{Bonora:2011gz} we analyzed the general consequences of adding one such gCS term to the action, in particular the appearance of a new contribution to the thermodynamical entropy. In \cite{BCDPSp2} we considered the global geometrical aspects implied by the presence of a gCS term, both at the level of the action and the entropy, and studied the topological conditions for the well-definiteness of both. Except in the three-dimensional case, very well studied in the literature, it does not seem to be easy to see the effects of a gCS term on observables. In the simplest and more symmetric cases they appear to be null. For this reason in \cite{Bonora:2012eb} we studied the case of Myers-Perry black hole in seven dimensions. We were able, at least perturbatively, to show that in some sufficiently complicated configuration the effects of the gCS are not identically vanishing. 

In this paper we would like to enlarge the null effect results of \cite{Bonora:2011mf}, with the purpose of circumscribing as closely as possible the cases in which the addition of a gCS
is irrelevant from an observational point of view. More to the point
we are interested in the gravity theories in $D=2n-1$ dimensions ($n\in2\mathbb{N}$) with 
Lagrangians of the form
\begin{equation} \label{lagrgen}
\LL = \LL_0 + \lambda \, \LL_{\mathrm{gCS}}
\end{equation}
where $\LL_0$ is some general manifestly diffeomorphism-invariant Lagrangian density
and $\LL_{\mathrm{gCS}}$ is the gCS Lagrangian density. In (\ref{lagrgen}) $\lambda$ denotes the gCS coupling constant. It is dimensionless and may be quantized, see \cite{BCDPSp2,Witten:2007kt,Lu:2010sj}. gCS terms have a remarkable set of properties, among which the most notable are: they are not manifestly diff-covariant, though they preserve diff-covariance in the bulk; they have a topological nature which leads to a quantization of their coupling; they are parity-odd and so break parity symmetry; they are conformally covariant, in the sense that under a Weyl rescaling of the metric
\begin{equation} \label{weylg}
\tilde{g}_{\mu\nu}(x) = \Omega^2(x) \, g_{\mu\nu}(x)
\end{equation}
Chern-Simons density transforms as (see \cite{CS,Solodukhin:2005ns} and Appendix),
\begin{equation} \label{weylL}
\LL_{\mathrm{gCS}}[\widetilde{\Gam}] = \LL_{\mathrm{gCS}}[\Gam] + d \,(\ldots)
\end{equation}
It is clear that gCS Lagrangian terms have a peculiar role in the set of all possible higher-curvature gravity terms, which makes them deserve special attention.

In the following we shall explicitly refer mainly to irreducible gCS terms, whose Lagrangian density is
given by
\begin{equation} \label{LgCS}
\LL_{\mathrm{gCS}}[\Gam] = n \int_0^1 dt \, \mathrm{str} (\Gam \, \R_t^{n-1})
\end{equation}
Here $\R_t = t d\Gam + t^2 \Gam\Gam$, $\Gam$ is the Levi--Civita connection and $\mathrm{str}$ denotes a symmetrized trace, which is an irreducible invariant symmetric polynomial of the Lie algebra of the $SO(1,D-1)$ group, and all products are wedge products. A general gCS term is a linear combination of irreducible and reducible terms, where the form of the latter is obtained from $(D+1)$-dimensional relation
\begin{equation} \label{LgCSr}
d\, \LL_{\mathrm{gCS}} = \tr(\R^{m_1}) \ldots \tr(\R^{m_k}) \;, \qquad\quad
  2 \sum_{j=1}^k m_j = D+1 \;, \qquad m_j \in 2\mathbb{N}
\end{equation}
with $k>1$ ($k=1$ gives irreducible gCS term). For example, in $D=7$ aside from the irreducible there is also a reducible gCS term.\footnote{In string theories compactified to $D=7$, they appear in combination when gCS terms are present.}  We shall state in what way the obtained results extend to reducible gCS terms. 

As anticipated above, in this paper we want to improve on the results found in \cite{Bonora:2011mf}. Our aim is to identify the class of metrics for which the inclusion of a gCS term in (\ref{lagrgen}) does not affect solutions and corresponding physical quantities. We show that for a large class of solutions, the effect of a gCS Lagrangian term is in fact null, and solutions corresponding to the Lagrangian $\LL_0$ are also solutions corresponding to (\ref{lagrgen}). In the case in which such solutions describe black holes (of general horizon topology) we shall show that the black hole entropy is also unchanged. As the case $n=2$ ($D=3$) has already been studied in detail in literature (see, e.g., \cite{DJT1,DJT2,Solodukhin:2005ah,Kraus:2006wn,Li:2008dq}), we focus here on $n\ge4$ ($D\ge7$).\footnote{In string theory $n=2$ gravitational CS terms play an important and unique role in some black hole analyses (see, e.g., 
\cite{Kraus:2006wn,Sen:2007qy,Cvitan:2007hu,Prester:2008iu,deWit:2009de}).} A particularly important intermediate result is the remark that the terms representing the gCS contribution to the entropy is invariant under local rescaling of the metric.

The paper is organized as follows. In section 2 we state and prove three theorems on the vanishing properties of the generalized Cotton tensor, the Weyl invariance of the gCS entropy and the vanishing of the latter under some general conditions. In section 3 we apply such theorems to various physical situations and in section 4 to linearized equations of motion around some highly symmetrical backgrounds.

\section{Three theorems}
\label{sec:theorems}

The paper is based on three results that we state in the form of theorems. The first
concern the effects of a gCS term on the generalized Cotton tensor. The second the invariance of the gCS entropy contribution under Weyl rescalings of the metric. Thanks to these result the third states that metrics such as those in the first theorem do not contribute to the gCS entropy. 

\subsection{Equations of motion}
\label{ssec:eom}

Adding a gCS term in the Lagrangian brings about additional terms in the equations of motion. 
It was shown in \cite{Solodukhin:2005ns} that the equation for the metric tensor $g_{\alpha\beta}$ 
acquires an additional term $C^{\alpha\beta}$, which, for the irreducible gCS term (\ref{LgCS}), is of the form
\begin{equation}\label{gCSeom}
C^{\alpha\beta} = - \frac{1}{2} \ \epsilon^{\mu_1 \cdots \mu_{2n-2} (\alpha} \, \nabla_{\!\rho}  \, \left( \tensor{R}{^{\beta)}_{\sigma_1}_{\mu_1}_{\mu_2}} \, 
\tensor{R}{^{\sigma_1}_{ \sigma_2 }_{\mu_3}_{\mu_4}} \cdots 
\tensor{R}{^{\sigma_{n-3}}_{ \sigma_{n-2}}_{\mu_{2n-5}}_{\mu_{2n-4}}}
\tensor{R}{^{\sigma_{n-2}}^{\rho}_{\mu_{2n-3}}_{\mu_{2n-2}}} \right)
\end{equation}
Under the Weyl rescaling of metric (\ref{weylg}) the tensor $C^{\alpha\beta}$ transforms as
\begin{equation} \label{Cconf}
C^{\alpha\beta}[\tilde{g}] = \Omega^{-(2D+2)} \, C^{\alpha\beta}[g] 
\end{equation}
Aside from being conformally covariant, the tensor $C^{\alpha\beta}$ is also traceless and covariantly conserved and so may be considered as a generalization of the Cotton tensor to $D=4k-1$ dimensions for $k>1$ \cite{Solodukhin:2005ns}.

It has been shown in the literature that due to their special symmetry properties, gCS contributions to equations of motion (\ref{gCSeom}) vanish for whole classes of metrics, such as maximally symmetric spaces (and conformally connected metrics) \cite{Solodukhin:2005ns}, and spherically symmetric metrics (with $SO(D-1)$ isometry subgroup) \cite{Bonora:2011mf,Lu:2010sj}. Here we want to show that there is a much broader class of metrics for which tensor $C^{\alpha\beta}$ vanishes.  This is guaranteed by the following theorem.

\bigskip

\noindent
\textbf{Theorem 1.} Assume that the metric of $D$-dimensional spacetime $(M,g_{\mu\nu})$ can be cast, in some region 
$\mathcal{O} \subset M$, in the following form,
\be \label{gconj}
ds^2 = g_{\mu\nu}(x)\,dx^\mu dx^\nu
 = D(x) \left( A(z)\,g_{ab}(y)\,dy^a dy^b + B(y)\,h_{ij}(z)\,dz^i dz^j \right) \ ,
\ee
where local coordinates on $\mathcal{O}$ are split into $x^\mu = (y^a, z^i)$, $\mu \in \{1,\ldots,D\}$, $a \in \{1,\ldots,p\}$, and $i \in \{1,\ldots,q\}$ (so that $p+q=D$).  The functions $B(y)$, $g_{ab}(y)$ and $A(z)$, 
$h_{ij}(z)$ depend only on the $\{y^a\}$ and $\{z^i\}$ coordinates, respectively. If $p \ge 2$ and $q \ge 2$ then for all $x \in \mathcal{O}$
\begin{equation} \label{Cgconj}
C^{\mu\nu}(x) = 0
\end{equation}

\bigskip

\noindent
\emph{Proof.} Due to property (\ref{Cconf}), equality (\ref{Cgconj}) is preserved under Weyl rescalings 
(\ref{weylg}). By taking $\Omega = (DAB)^{-1}$ the metric (\ref{gconj}) may be put in the
direct product form
\begin{equation} \label{gblock}
d\tilde{s}^2 = \tilde{g}_{\mu\nu}\,dx^\mu dx^\nu = g_{ab}(y)\,dy^a dy^b + h_{ij}(z)\,dz^i dz^j \ ,
\end{equation}
so we only have to prove that the theorem hold for the metrics of the type (\ref{gblock}). This greatly simplifies our job because both Riemann tensor and its covariant derivative are completely block-diagonal, and as a consequence also the tensor
\begin{equation} \label{Cpart}
\nabla_{\!\rho}  \, \left( \tensor{R}{^{\beta}_{\sigma_1}_{\mu_1}_{\mu_2}} \, 
\tensor{R}{^{\sigma_1}_{ \sigma_2 }_{\mu_3}_{\mu_4}} \cdots 
\tensor{R}{^{\sigma_{n-3}}_{ \sigma_{n-2}}_{\mu_{2n-5}}_{\mu_{2n-4}}}
\tensor{R}{^{\sigma_{n-2}}^{\rho}_{\mu_{2n-3}}_{\mu_{2n-2}}} \right)
\end{equation}
present in the definition of $C^{\mu\nu}$ (\ref{gCSeom}) is. This means that the components of the tensor in
(\ref{Cpart}) are nonvanishing only when all the indices are either from the $y$-subspace or the $z$-subspace. Because there are $D-1$ free indices in (\ref{Cpart}) which are contracted with the totally antisymmetric Levi-Civita tensor in (\ref{gCSeom}) (that is, all have to be mutually different) it is obvious that when both $p>1$ and $q>1$ then (\ref{Cpart}) is zero, implying that $C^{\mu\nu}$ is also zero. $\blacksquare$

\medskip

For reducible gCS terms the Theorem 1 gets modified, allowing other possibilities (aside $p$ or $q$ equal to 0 or 1) in which one may have $C^{\alpha\beta} \ne 0$ for geometries of the type (\ref{gconj}). Their contribution to $C^{\alpha\beta}$ is a sum of terms which are of the form \cite{Bonora:2011mf}
\begin{equation}
\epsilon^{\mu_1\cdots\mu_{D-1}(\alpha}
\left( \tr(\R^{2m_1}) \cdots \tr(\R^{2m_{k-1}}) \right)_{\mu_1 \cdots \mu_{D+1-2m_k}}
\nabla_{\!\rho}  \, \left( \tensor{R}{^{\beta)}_{\sigma_1}_{\mu_{D+2-2m_k}}_{\mu_{D+3-2m_k}}} \cdots 
\tensor{R}{^{\sigma_{n-2}}^{\rho}_{\mu_{D-2}}_{\mu_{D-1}}} \right)
\end{equation}
Following the same logic as above it is easy to conclude that for the reducible gCS term, defined implicitly by (\ref{LgCSr}), exceptions to Theorem 1 may appear when there is a subset $\{m_{j_1},\ldots,m_{j_l}\}$ of the set of exponents $\{m_{j}, j=1,\ldots,k\}$ such that
\begin{equation} \label{T1red}
2\sum_{r=1}^l m_{j_r} + \sigma \,=\, p \;\mbox{ or }\; q 
\end{equation}
for $\sigma=0$ or 1. When $p$ or $q$ satisfy (\ref{T1red}) it is possible that $C^{\alpha\beta} \ne 0$. For example, in $D=7$ there is a unique reducible gCS term, which has $k=2$ and $m_1=m_2=2$, so (\ref{T1red}) gives no restrictions on $p$ and $q$ (all values from 0 to 7 are allowed), so in this case Theorem 1 by itself has no content. However, if any of the submetrics, $g_{ab}$ or $h_{ij}$, is maximally symmetric and $p>1$ or $q>1$, one has\footnote{This follows because maximally symmetric spaces satisfy (\ref{msprop}).} $C^{\alpha\beta} = 0$ and so in this case the original statement of the Theorem applies also to all reducible gCS terms. This will be relevant when we discuss applications in 
Sec. \ref{sec:examples}

The obvious consequence of the above theorem is that if we can find coordinates around every point of spacetime in which the metric, which is a solution to the equations of motion obtained from some Lagrangian $\LL_0$, is of the form (\ref{gconj}), then this metric will also be a solution in the theory defined by the Lagrangian (\ref{lagrgen}). In other words, adding gCS Lagrangian terms does not affect solutions which are of the form specified by the theorem. 

The theorem covers many classes of metrics frequently discussed in the literature. In Sec. \ref{sec:examples} we shall mention a few examples of particular interest.

\subsection{Black hole entropy}
\label{ssec:entropy}

If the metric describes a black hole one is also interested in its thermodynamical behavior, and in particular in the black hole entropy. It was shown in \cite{Tachikawa:2006sz} that the irreducible gCS Lagrangian term (\ref{LgCS}) brings an additional term in the black hole entropy formula, which must be added to Wald's formula \cite{Iyer:1994ys} obtained from the $\LL_0$ part of the total Lagrangian 
(\ref{lagrgen}), given by \cite{Bonora:2011gz}
\begin{equation} \label{entgCS}
S_{\mathrm{gCS}}[g] =  4\pi n \int_\mathcal{B} \boldsymbol{\omega} (d\boldsymbol{\omega})^{n-2} \ ,
\end{equation}
where $\mathcal{B}$ is the $(D-2)$-dimensional bifurcation surface of the black hole horizon 
and $\boldsymbol{\omega}$ is a 1-form to be identified with the $SO(1,1)$ (or $U(1)$ in Euclidean signature) connection on the normal bundle of $\mathcal{B}$. Here we want to discuss some general properties of the gCS entropy term (\ref{entgCS}).

\bigskip
\noindent
\textbf{Theorem 2.} The gCS entropy term (\ref{entgCS}) is invariant under Weyl rescalings (\ref{weylg}) of the spacetime metric. 

\bigskip

\noindent
\emph{Proof.} In \cite{Bonora:2011gz} we showed that $\boldsymbol{\omega}$ can be written as
\begin{equation}
\omega_\mu = -\tensor{q}{_\mu^\nu}\,n_\rho \nabla_{\!\nu}\,\ell^\rho
\end{equation}
where $q_{\mu\nu}$ is the induced metric on $\mathcal{B}$, and $\ell^\mu$ and $n^\mu$ is a pair of two future directed null vector fields, normal to the black horizon and arbitrary up to a normalization $\ell^\mu n_\mu = -1$. For the Weyl rescaled metric (\ref{weylg}) we can take the null vectors to be
\begin{equation}
\tilde{\ell}_\mu = \sqrt{\Omega}\, \ell_\mu \ , \qquad \tilde{n}_\mu = \sqrt{\Omega}\, n_\mu
\end{equation}
By using $\tilde{q}_\mu{}^\nu = q_\mu{}^\nu$ and a well-known relation (e.g., see Appendix D of \cite{Wald:1984rg})
\begin{equation}
\widetilde{\nabla}_{\!\nu} \ell^\rho = \nabla_{\!\nu} \ell^\rho + C^\rho_{\alpha \nu} \ell^\alpha \ , \qquad
 C^\rho_{\alpha \nu} = \delta^\rho_{(\alpha} \nabla_{\!\nu)} \ln \Omega
  - \frac{1}{2} g_{\alpha \nu} g^{\rho \beta} \nabla_{\!\beta} \ln \Omega
\end{equation}
a straightforward calculation gives
\begin{equation}
\widetilde{\omega}_\mu = \omega_\mu \ .
\end{equation}
Using this in (\ref{entgCS}) we obtain
\begin{equation}
S_{\mathrm{gCS}}[\tilde{g}] = S_{\mathrm{gCS}}[g]
\end{equation}
which proves the theorem. 
$\blacksquare$

\medskip

\subsection{Vanishing of gCS entropy}
\label{ssec:vangCS}

The third theorem describes some general consequences of the first two.

\bigskip
\noindent
\textbf{Theorem 3.} If a metric $g_{\mu\nu}(x)$ describing a black hole is of the form (\ref{gconj}), with 
$q \ge 1$ and where the coordinates $z$ are tangential to the bifurcation surface of the horizon, then the gCS entropy term (\ref{entgCS}) evaluated on such metric vanishes
\begin{equation}
S_{\mathrm{gCS}}[g] = 0 \ .
\end{equation}

\bigskip

\noindent
\emph{Proof.} First we make a Weyl rescaling (\ref{weylg}) with $\Omega = (DAB)^{-1}$ to obtain the metric
$\tilde{g}_{\mu\nu}$ in the direct product form (\ref{gblock}). Theorem 2 says that the gCS entropy term is invariant under such transformation, so we can use $\tilde{g}_{\mu\nu}$ to evaluate it. Let us focus on
the components of $\omega_\mu$ in ``$z$-directions", i.e., for $\mu = i$. Due to the block diagonality of the metric we have
\begin{equation}
\widetilde{\omega}_i = -\tensor{\tilde{q}}{_i^j}\,\tilde{n}_a \widetilde{\nabla}_{\!j}\,\tilde{\ell}^a
\end{equation}
where we have used the fact that metric $\tilde{g}_{\mu\nu}$ has direct product form and that $\tilde{n}$ and $\tilde{\ell}$ are defined purely in the $y$-block. Moreover, the direct product form of the metric implies that the covariant derivative $\widetilde{\nabla}_{\!j}$ is defined purely in the $z$-block. From this follows that 
$\widetilde{\nabla}_{\!j}\,\tilde{\ell}^\rho = 0$ and so $\widetilde{\omega}_i = 0$. As indices from the $z$-block must appear when performing the integration in (\ref{entgCS}), it directly follows that
\begin{equation}
S_{\mathrm{gCS}}[\tilde{g}] = 0, 
\end{equation}
which completes the proof of the theorem
$\blacksquare$

\medskip

Using the results from \cite{Bonora:2011gz} (see Eq. (4.26)) and conformal properties of $\tr(\R^k)$ 
\cite{CS,Solodukhin:2005ns} it directly follows that the theorems 2 and 3 are valid also for the reducible gCS terms.

\vspace{10pt}

\section{Applications}
\label{sec:examples}

\bigskip

The Theorem 1 from Sec. \ref{ssec:eom} covers many classes of metrics appearing in different contexts. Here we discuss a few situations frequently occurring in the literature.

\subsection{Spacetimes with maximally symmetric subspaces}
\label{ssec:maxsym}

If the spacetime allows a foliation with maximally symmetric $d$-dimensional subspaces (with $d \ge 2$), then it is known that one can find coordinates in which the metric has the form (\ref{gconj}) with 
$D(x) = A(z) = 1$ and $q = d$ (see, e.g., Sec. 13.5 of \cite{Weinberg}). There are at least three frequent contexts where such metrics appear:
\begin{enumerate}
\item \emph{Cosmology}
-- If the metric describes a cosmological model of the Universe (in some extra-dimensional set-up) then the $SO(3)$ isometry subgroup, following from the isotropy of 3-dimensional ``physical" space, implies that the spacetime may be foliated by 2-spheres (in this case we have $d=2$). 
\item \emph{Stationary rotating black holes}
-- If a stationary rotating black hole has $k$ angular momenta vanishing with $k\ge2$, this typically implies that the isometry group has an $SO(2k)$ factor. Then there exists a foliation in $(2k-1)$-spheres, so here $d = 2k-1$. In this case Theorems 1 and 3 apply. In the $k=1$ case it may naively seem that Theorem 3 implies the contribution of the gCS entropy term again to be vanishing; however this is not so. We have explicitly shown in \cite{Bonora:2012eb} on a particular example in $D=7$ (which gives Myers-Perry solution \cite{Myers:1986un} for $\lambda=0$) that the effect of the gCS term in the equations of motion in such a case is such that it forces the metric to depart from the form (\ref{gconj}). So we cannot apply Theorem 3 to the full black hole solution.
\item \emph{Flat $p$-branes}
-- Geometries of flat $p$-branes are of the form (\ref{gconj}) with $q=p+1$, so for $p \ge 1$ Theorem 1 applies. If in addition they are black $p$-branes, then also Theorem 3 applies. 
\end{enumerate}

\subsection{Scenarios in extradimensional theories}
\label{ssec:extdim}

In many extra-dimensional scenarios appearing more or less frequently in the literature, either in the form of braneworlds or Kaluza-Klein (KK) compactifications, the metric of the vacuum is of the form (\ref{gconj}). In realistic scenarios $p=4$, thus it follows that $q = D-p \ge 3$ (because the theories we consider in this paper have $D\ge7$); so Theorem 1 applies. Notice that it applies also for any perturbation of such vacuum provided the metric is still of the form (\ref{gconj}) with $q\ge2$. Thus a gCS Lagrangian term does not affect solutions of this form. 

Let us explain the situation with a simple example. We consider some diff-covariant theory in $D=11$ and add to it an $n=6$ gCS Lagrangian term (\ref{LgCS}). Then we proceed to the standard KK reduction to $D=4$. If in the vacuum all the seven KK gauge fields coming from the metric vanish, then the vacuum metric is of the form (\ref{gconj}), with $m=4$ and $q=7$. If we excite the vacuum by switching on (among other) $k$ gauge fields, then the metric will still be of the form (\ref{gconj}), but now with $q = 7-k$. This implies that for $k \le 5$ Theorem 1 still applies. In fact to be able to see a nontrivial effect of gCS Lagrangian term one needs to analyze configurations with at least five nonvanishing KK gauge fields (coming from the metric). That is, apart from the D=3 case, one needs fairly complicated configurations to be able to infer existence of pure gravitational Chern-Simons Lagrangian term coupled to a theory.

\vspace{10pt}

\section{Linearization around maximally symmetric backgrounds}
\label{sec:linear}

\bigskip

It is quite obvious that gCS Lagrangian terms do not contribute to the linearized equations of motion (EOM) around a flat Minkowski background metric. We now show that this also holds for more general backgrounds, including (A)dS metrics.

\bigskip
\noindent
\textbf{Theorem 4.} In $D>3$ (irreducible (\ref{LgCS}) and reducible) gCS Lagrangian terms do not affect the linearized equations of motion when the background space is maximally symmetric or is a product of maximally symmetric spaces.

\bigskip

\noindent
\emph{Proof.} Linearized EOM's around background metrics $g_{(0)\mu\nu}$ are obtained by writing the metric as
\begin{equation} \label{linmet}
g_{\mu\nu} = g^{(0)}_{\mu\nu} + h_{\mu\nu}
\end{equation}
and expanding the equations of motion around $g_{(0)\mu\nu}$ while keeping only the terms which are at most first-order in $h_{\mu\nu}$. We now show that inserting (\ref{linmet}) in  the $C_{\mu\nu}$ term 
(\ref{gCSeom}) and expanding in $h_{\mu\nu}$, does not produce terms of zeroth and first order whenever the background metric is maximally symmetric. First note that Theorem 1 applies to all maximally symmetric metrics, so there is no contribution at zeroth order
\begin{equation}
C^{(0)}_{\mu\nu} = C_{\mu\nu}[g^{(0)}] = 0
\end{equation} 
Let us turn next to first order terms. We use the following properties of maximally symmetric metrics
\begin{equation} \label{msprop}
\nabla^{(0)}_\mu R^{(0)}_{\nu\rho\sigma\kappa} = 0 \ , \qquad 
 \left( \R_{(0)}^2 \right)^\alpha{}_\beta \equiv 
 \R_{(0)}^\alpha{}_\gamma \wedge \R_{(0)}^\gamma{}_\beta = 0
\end{equation}
where $\R$ is the tensor-valued 2-form curvature defined by 
\begin{equation}
(\R^\alpha{}_\beta)_{\mu\nu} = R^\alpha{}_{\beta\mu\nu} \ .
\end{equation}
The second equation in (\ref{msprop}) follows from
\begin{equation} \label{msmet}
R^{(0)}_{\mu\nu\rho\sigma} = 
 \pm \ell^{-2} \left( g^{(0)}_{\mu\rho} g^{(0)}_{\nu\sigma} - g^{(0)}_{\mu\sigma} g^{(0)}_{\nu\rho} \right)
\end{equation}
and from maximally symmetric metrics being diagonal. It is obvious from the form of (\ref{gCSeom}) that relations
(\ref{msprop}) guarantee that there are no first order terms in $h_{\mu\nu}$  when $n \ge 6$ ($D \ge 11$). In $D=7$ ($n=4$) there is one suspicious term
\begin{equation} \label{Cfirst}
C_{(1)}^{\alpha\beta} = - \frac{1}{2} \epsilon_{(0)}^{\mu_1 \cdots \mu_6(\alpha}
R_{(0)}^{\beta)}{}_{\sigma_1\mu_1\mu_2} \nabla^{(0)}_\rho R_{(1)}^{\sigma_1}{}_{\sigma_2\mu_3\mu_4}
R_{(0)}^{\sigma_2\rho}{}_{\mu_5\mu_6}
\end{equation}
coming from the irreducible gCS term, which is not obviously vanishing. However, by using Eqs. (7.5.7)--(7.5.8) and (3.2.12) from \cite{Wald:1984rg} one can put (\ref{Cfirst}) in a form in which the second equation in (\ref{msprop}) and (\ref{msmet}) again force it to vanish.
If the background is a direct product of maximally symmetric spaces, the proof follows in the same way. The only difference is that in (\ref{msprop}) and (\ref{msmet}) there is a different radius $\ell_i$ for every maximally symmetric subspace $i$, but this does not affect the proof.
Note that the theorem is valid off-shell, i.e. regardless of whether the background metric satisfies the EOM or not.
$\blacksquare$

\medskip

There are many important situations where linearization enters. Let us mention three of them and emphasize direct consequences of Theorem 4: (1) Perturbative degrees of freedom around flat and (A)dS spaces in $D>3$ dimensions - their number and properties are unchanged after introducing a gCS Lagrangian term. If we add this term to the Hilbert-Einstein Lagrangian, we still have just one massless spin-2 excitation (graviton). (2) Stability analysis of a solution with metric 
$g_{(0)\mu\nu}$ - gCS Lagrangian terms do not affect the stability analysis of maximally symmetric spaces (or their products) in $D>3$. (3) Determination of asymptotic charges, in particular mass and angular momenta, which in the method described in Section 7.6 of \cite{Weinberg} (asymptotically flat configuration) or \cite{Deser:2002jk} (asymptotically AdS configurations) are calculated from linearized EOM's - in $D>3$ dimensions the $C_{\mu\nu}$ term does not contribute directly to charges, so that the only possible contribution of a gCS Lagrangian term is indirect through changing the asymptotic behavior of the metric (in the spatial infinity). In \cite{Bonora:2012eb} we have shown, on an explicit example of stationary rotating black hole in $D=7$, that the gCS term does not change the relevant asymptotic behavior of the metric and, as a consequence, relations for mass and angular momentum are perturbatively unchanged when a gCS Lagrangian term is introduced.\footnote{This was explicitly shown up to first order in gCS coupling $\lambda$, but, based on the structure of the gCS contribution to the EOM, we conjectured that the result is valid to all orders in $\lambda$.} 
It remains to be shown how general this result is, and in particular whether it is valid also for asymptotically (A)dS configurations.

It should be emphasized that the $n=2$ ($D=3$) case, which is excluded in Theorem 4, is indeed exceptional. It was shown in \cite{DJT1,DJT2} that in $D=3$ the gCS term contributes to the linearized EOM, in a way which may make graviton massive (like in Topologically Massive Gravity), 
and/or generates additional terms in the expressions for mass and angular momentum 
\cite{Deser:2003vh,Olmez:2005by}. As discussed above, Theorem 4 guarantees that nothing like this happens in $D>3$.

\vspace{10pt}

\section{Conclusion}
\label{sec:concl}

\bigskip

This paper is another step in our endeavor to understand the consequences and nature of adding a purely gravitational Chern-Simons (gCS) term to an otherwise ordinary gravitational action. With the exception of the D=3 case the effects of such addition are rather elusive, and are present only for configurations with rather modest space-time symmetries. Conversely the problem of circumscribing metric solutions to the equations of motion that are left unchanged by the same addition is also very elusive. Similarly far from obvious is the related question of whether the effects of gCS terms are of topological nature or not. We believe the best course in this situation is to try to enlarge as far as possible the class of cases where the effect of a gCS term are null. This is what we have done in this paper. We have proved three theorems that allow us to conclude for a rather large class of metrics that the corresponding physics is not affected by the addition of a gCS term. It include cosmological, black hole and p-brane solutions. We have also shown that these theorems are helpful for a much larger class of problems in which linearized gravity is involved.

\vspace{20pt}

\noindent
{\bf \large Acknowledgements}

\bigskip

\noindent
One of us (L.B.) would like to thank the Theoretical Physics Department, University of Zagreb, for hospitality and financial support during his visits there. The work of L.B. was supported in part by the MIUR-PRIN contract 2009-KHZKRX.. M.C., P.D.P., S.P.\ and I.S.\ would like to acknowledge support by the Croatian Ministry of Science, Education and Sport under the contract no.~119-0982930-1016.

\vspace{20pt}

\section*{Appendix}
\appendix

\bigskip

\section{A simple proof of the Cotton tensor's conformal covariance}
\label{app:conf-inv}

\bigskip

There is a simple way to derive the property (\ref{Cconf}), based on theorem by Chern and Simons 
\cite{CS}. It is instructive to review it because it highlights the global issues underlying the proof. Let $LM$ be the frame bundle over the manifold $M$ (with structure group
$GL(D)$) and let ${\boldsymbol \phi}$ be a connection on this bundle, ${\boldsymbol \phi}_t= t{\boldsymbol\phi} $ and ${\mathbf \Phi},{\mathbf \Phi}_t$ the respective curvatures. Let $P$ denote an invariant polynomial of $GL(D)$ and let us write, as usual, the transgression
\be\label{TPphi}
TP({\boldsymbol \phi})= n\,\int_0^1dt \,P( {\boldsymbol \phi}, {\mathbf \Phi}_t,\ldots,{\mathbf \Phi}_t) 
\ee
The theorem says:

\textbf{Theorem (CS).}
If $g$ and $\tilde g$ are conformally related, (\ref{weylg}), and ${\boldsymbol \phi},{\mathbf \Phi}$ and $\tilde {\boldsymbol \phi},\tilde {\mathbf \Phi}$ are
their respective connections and curvatures, we have
\begin{displaymath}
TP(\tilde {\boldsymbol \phi})=TP({\boldsymbol \phi})+ d {\mathbf \Theta}^{\mathrm{ncf}}
\end{displaymath}

In the case ${\boldsymbol\phi}$ is a Riemannian connection, $TP({\boldsymbol \phi})$ reduces to $\CS_{\mathrm{CS}}(\Gam)$, see \cite{Bonora:2011mf}, that is to gCS Lagrangian term.
Now, the Cotton tensor is defined via the relation
\be \label{deltaCS}
\delta \CS_{\mathrm{CS}}(\Gamma) = C^{\mu\nu}\delta g_{\mu\nu} \, {\boldsymbol \epsilon}  + d{\Th}^{\mathrm{cov}}
 + d{\Th}^{\mathrm{nc}}
\ee
where ${\boldsymbol \epsilon}= \sqrt{-g}d^Dx$. Taking the analogous variation for
$\CS_{\mathrm{CS}}(\Gamma)$
\be \label{deltaCS'}
\delta \CS_{\mathrm{CS}}(\tilde \Gamma) = \tilde C^{\mu\nu}\delta \tilde g_{\mu\nu} \,  
 \tilde{\boldsymbol \epsilon} + d \tilde{\Th}^{\mathrm{cov}} + d \tilde{\Th}^{\mathrm{nc}}
\ee
where $\delta \tilde g= \Omega^2 \delta g$, 
comparing the two and taking into account the transformation properties of the volume element, one gets immediately (\ref{Cconf}).

In deriving the equations of motion any exact term in the previous formulas is discarded, but in the derivation of the entropy formula by means of the phase space formalism also ${\Th}^{\mathrm{cov}}$ and ${\Th}^{\mathrm{nc}}$ play a role. Therefore  the term ${\mathbf \Theta}^{\mathrm{ncf}}$ has to be taken into account when comparing conformally related metrics. As a consequence it is not a priori obvious that the entropy formula for a gCS term is conformally invariant. But this turns out to be the case.

\vspace{30pt}


\end{document}